\documentclass[preprint,12pt]{elsarticle}

\usepackage{amssymb}

\usepackage[export]{adjustbox}
\usepackage{subcaption}
\usepackage{caption}
\usepackage{lineno}
\usepackage{setspace} 
\usepackage{graphicx}
\usepackage{chngcntr}
\usepackage{titlesec}

\usepackage{amsmath}
\journal{Acta Materialia}

\begin{document}

\begin{frontmatter}



\title{A cascade model for the defect-driven etching of porous GaN distributed Bragg reflectors}


\author[Cambridge,First Author,Corresponding Author]{Ben Thornley} %% Author name
\author[Cambridge,First Author]{Maruf Sarkar}
\author[Swansea]{Saptarsi Ghosh}
\author[Cambridge]{Martin Frentrup}
\author[Cambridge]{Menno J. Kappers}
\author[Cambridge]{Thom R. Harris-Lee}
\author[Cambridge]{Rachel A. Oliver}

\affiliation[Cambridge]{organization={Department of Materials Science and Metallurgy, University of Cambridge},%Department and Organization
            addressline={27 Charles Babbage Road}, 
            city={Cambridge},
            postcode={CB3 0FS}, 
            country={United Kingdom}}

\affiliation[Swansea]{organization={Department of Electronic and Electrical Engineering, Swansea University},%Department and Organization
            addressline={Swansea University Bay Campus, Fabian Way}, 
            city={Swansea},
            postcode={SA1 8EN}, 
            country={United Kingdom}}

\affiliation[First Author]{country={These authors contributed equally to this work}}

\affiliation[Corresponding Author]{country={{Corresponding author - Contact: bdt28@cam.ac.uk +44 (0)1223 3 31954
}}}

%% Abstract
\begin{abstract}

Fabrication of porous GaN distributed Bragg reflectors (DBRs) via the selective electrochemical etching of conductive Si-doped layers, separated by non-intentionally doped (NID) layers, provides a straightforward methodology for producing highly reflective DBRs suitable for device overgrowth and integration, which has otherwise proven difficult in the III-nitride epitaxial system via conventional alloying. Such photonic materials can be fabricated by a lithography-free defect-driven etching process, where threading dislocations intrinsic to heteroepitaxy form nanoscale channels that facilitate etchant transport through NID layers. Here, we report the first three-dimensional characterisation of porous GaN-on-Si DBRs fabricated in this methodology with different etching voltages, using serial-section tomography in a focused ion beam scanning electron microscope (FIB-SEM). These datasets reconstruct the pore morphology as etching proliferates through the alternating Si-doped/NID layer stack. Volumetric reconstruction enabled us to enhance the established `kebab' model for defect-driven etching by proposing a `cascade' model where the etchant cascades through the material via vertical etching down nanopipes and horizontal etching across pores, forming complex networks directly related to the pathways taken. This accounts for premature nanopipe termination and discontinuities in nanopipe formation, where dislocations are observed to activate and deactivate individually. Statistical analysis of individual etching behaviour, across all dislocations for each tomograph, revealed a greater tendency to form continuous structures that follow conventional `kebab' behaviour at higher etching voltages. We propose that higher etching voltages alter the probability of dislocation etching relative to doped layer etching, thereby empowering morphological optimization through improved mechanistic understanding of electrochemical etching. 

\end{abstract}

\begin{graphicalabstract}
\includegraphics[width=1\linewidth]{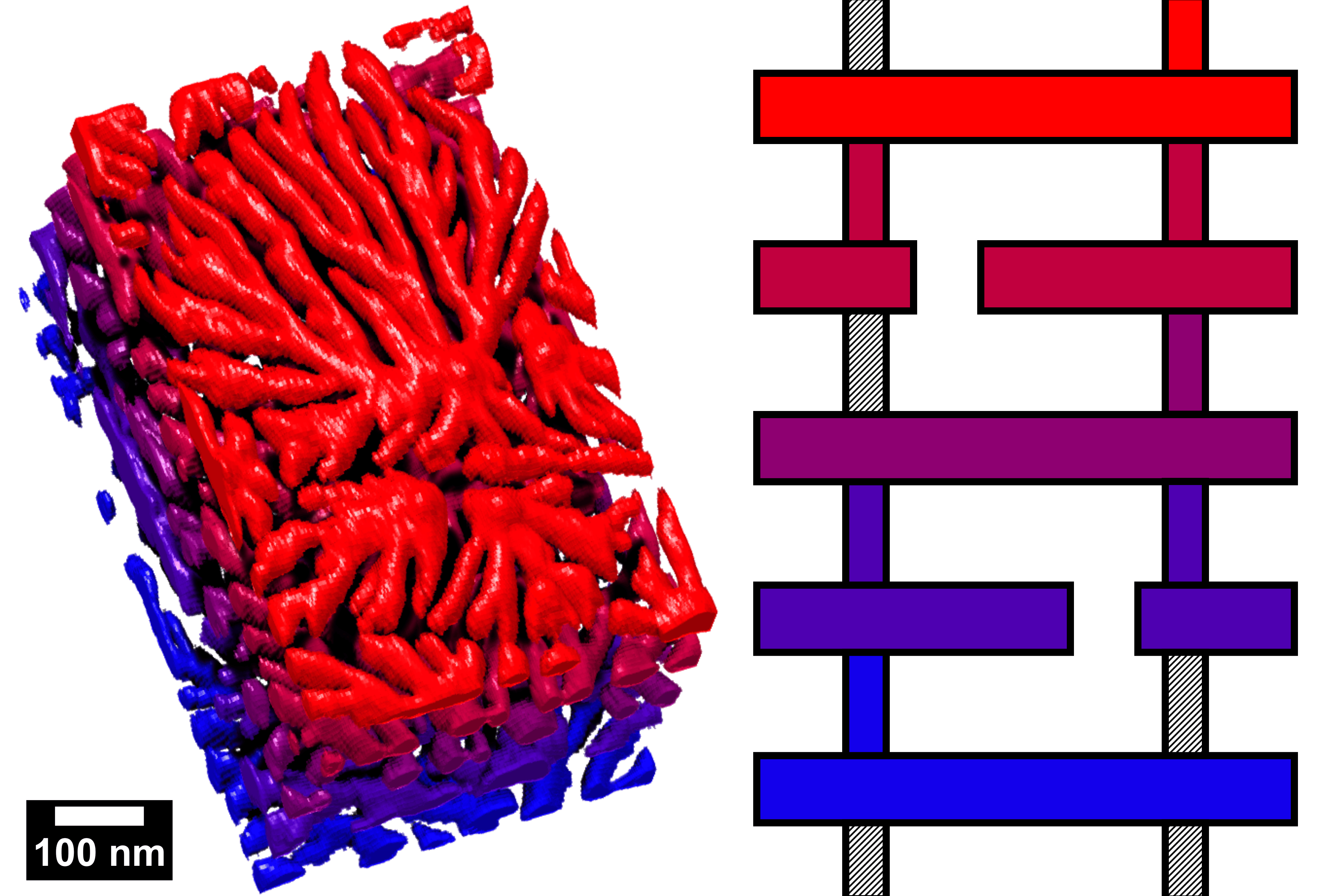}
\end{graphicalabstract}


%% Keywords

\end{frontmatter}



\section{Introduction}
\label{Introduction}

The introduction of porosity into III-nitride semiconductors via a conductivity-selective electrochemical etching process allows for a new dimension for the engineering of material properties \cite{Griffin_Oliver_2020}. This porosification process is performed using a simple electrolytic cell, where a gallium nitride (GaN) sample and a connected platinum electrode are submerged in a conductive electrolyte solution (`etchant') with a voltage applied across them. Prominent examples of properties which may be controlled precisely through varying fractional porosity are the extent of strain relaxation of epilayers overgrown onto porous substrates \cite{Yang_Chen_Cao_Zhao_Shen_Luan_Pang_Liu_Ma_Xiao_2019}, the piezoelectric coefficient of the GaN matrix \cite{CALAHORRA2020100858}, and the refractive index \cite{Zhang_Park_Chen_Lin_Xiong_Kuo_Lin_Cao_Han_2015}. Accordingly, this porosification technology can be said to offer an additional non-compositional degree of freedom in device engineering.

In this study, we investigate porous GaN distributed Bragg reflectors (DBRs). These are highly reflective, wavelength-selective mirrors that can be used to create a resonant cavity in optoelectronic devices \cite{Yao_Liang_Guo_Xiu_2023}, such as vertical cavity surface-emitting lasers \cite{10697339}, single photon sources \cite{10.1063/5.0049488}, and resonant cavity LEDs \cite{app11010008}. Porous GaN DBRs are fabricated through metalorganic chemical vapour deposition (MOCVD) to prepare an epilayer stack, followed by electrochemical etching to introduce nanoscale porosity \cite{mays_2007}. Here, a latent DBR multilayer stack is first grown by epitaxy of alternating layers of highly Si-doped GaN and non-intentionally doped (NID) GaN. Electrochemical etching then partially removes material in the highly Si-doped GaN, leaving porous GaN which acts as an effective medium with a reduced refractive index, whilst the NID GaN layers are largely unaffected and remain notionally non-porous \cite{zhu_liu_ding_fu_jarman_ren_kumar_oliver_2017}. This procedure thus results in a periodic porous/non-porous multilayer stack with alternating low/high refractive index, providing a material framework for the preparation of highly reflective DBRs.

Highly reflective DBRs require materials with substantial refractive index contrast between the alternating layers. Obtaining highly reflective structures using the conventional approach of compositional variation between alternating layers has proven less successful for III-nitrides than in other epitaxy systems; pairing AlN/GaN proves unsuccessful due to poor refractive index contrast ($\frac{\Delta n}{n} = 0.16$) and substantial lattice mismatch (2.4\,\%), resulting in significant challenges for strain management and crack prevention \cite{Yagi_Kaga_Yamashita_Takeda_Iwaya_Takeuchi_Kamiyama_Amano_Akasaki_2012}. Introduction of tertiary alloying elements can permit lattice matching, such as with Al$_{0.82}$In$_{0.18}$N/GaN, but this pair has an even lower refractive index contrast ($\frac{\Delta n}{n} = 0.06$) \cite{Carlin_Zellweger_Dorsaz_Nicolay_Christmann_Feltin_Butté_Grandjean_2005}. Furthermore, the epitaxial growth of AlInN is sufficiently slow as to be undesirable for mass adoption \cite{Jarman_Zhu_Griffin_Oliver_2019}. When designing with low refractive index contrast materials, DBR reflectance can be improved by increasing the number of reflecting layer pairs. However, introducing this additional epitaxial complexity can be costly or impose limits on device design \cite{Ghosh_Sarkar_Frentrup_Kappers_Oliver_2024}. In contrast, porous GaN DBRs offer high refractive index contrast between alternating layers and minimal lattice strain, therefore achieving stable structures which require fewer reflecting pairs, and preserve surface quality for high reflectance mirrors suitable for device overgrowth \cite{Jarman_Zhu_Griffin_Oliver_2019, Ji_Frentrup_Zhang_Pongrácz_Fairclough_Liu_Zhu_Oliver_2023}.

The first demonstration of the GaN/porous GaN DBR framework was the work of Zhang \textit{et al.} in 2015 \cite{Zhang_Park_Chen_Lin_Xiong_Kuo_Lin_Cao_Han_2015}. In their process, a thin silicon dioxide dielectric layer was deposited onto the DBR surface, and deep trenches defined by lithographic patterning techniques were created to allow physical contact between the liquid etchant and highly Si-doped GaN layers, to permit the dissolution of material and therefore enable porosification to progress throughout the doped layers \cite{Mishkat-Ul-Masabih_Luk_Rishinaramangalam_Monavarian_Nami_Feezell_2018}. Electrochemical etching, therefore, proceeded laterally from the exposed surfaces through the doped layers and results in aligned horizontal pores \cite{griffin_patel_zhu_langford_kamboj_ritchie_oliver_2020}. This work demonstrated the feasibility of the material framework, but the requirement of lithographic processes, which are complex and costly, restricted device design versatility and scalability.

Zhu \textit{et al.} developed an approach for fabricating porous GaN DBRs in the GaN/sapphire epitaxial system that circumvents the need for any lithography, whereby the electrochemical etching process proceeds via nanoscale vertical transport channels (or `nanopipes') formed by the local etching of the threading dislocations that arise in the highly mismatched heteroepitaxy of GaN on unlike substrates such as silicon and sapphire \cite{zhu_liu_ding_fu_jarman_ren_kumar_oliver_2017}. The formation of nanopipes on threading dislocations was demonstrated by TEM-based Burgers circuit analysis by Massabuau \textit{et al.} in 2020 \cite{massabuau_griffin_springbett_liu_kumar_zhu_oliver_2020}. These nanopipes are sufficient to allow etchant to move through the structure and contact the doped layers, permitting dissolution via electrochemical etching (and therefore porosification) whilst remaining small enough to ensure that NID layers are minimally affected and remain almost entirely non-porous. This approach allows for simple wafer-scale etching and leaves the surface suitable for further epitaxy and the overgrowth of device structures \cite{Jarman_Zhu_Griffin_Oliver_2019}.

Previous work regarding porous GaN-on-sapphire DBRs etched in a defect-driven manner initially suggested that all threading dislocations participate in electrochemical etching \cite{griffin_patel_zhu_langford_kamboj_ritchie_oliver_2020}. This simple model describes each threading dislocation being etched to form a nanopipe that runs through the entire multilayer stack, with pores emanating outward from the nanopipe at each highly Si-doped layer, which do not coalesce or merge \cite{massabuau_griffin_springbett_liu_kumar_zhu_oliver_2020}. The structure of individual pores under this framework is therefore characterised by a single central vertical nanopipe through the stack, with large porous fields emanating radially in all of the doped layers in the material. Throughout, we will refer to this model as the `kebab model', invoking the analogy of a threading dislocation as a shish-kebab skewer \cite{Wang_Chen_Zhang_Fu_2008}. We recently demonstrated that the defect-driven etching methodology first developed for GaN/sapphire also applies to the mass-market GaN-on-Si platform \cite{Ghosh_Sarkar_Frentrup_Kappers_Oliver_2024}. This work provided the first evidence that not all threading dislocations in GaN-on-Si epitaxy appear active (i.e., participate in electrochemical porosification to form a nanopipe with an associated field of porosity) during the etching process, thus suggesting that the simple kebab model may not be a complete framework.

Literature concerning pore morphology visualization and quantification in porous GaN has generally been restricted to cross-sectional secondary electron (SE) imaging of cleaved porous samples in the scanning electron microscope (SEM), which provides a limited view along the growth direction \cite{Bao_Lu_Wang_Liu_Han_Bi_Li_Huang_Kang_Kamiyama_et_al_2025}. However, the cross-sectional perspective can be misleading in revealing the extended morphology of the porous network \cite{yang_xiao_cao_zhao_shen_ma_2018}. For porous GaN DBRs, recent work has developed a novel sub-surface imaging modality based on the backscattered electron (BSE) signal as generated by a high 20\,keV primary electron landing energy in the SEM, to offer more insightful plan-view characterisation of the in-layer pore morphologies \cite{Sarkar_Adams_Dar_Penn_Ji_Gundimeda_Zhu_Liu_Hirshy_Massabuau_et_al_2024}. This non-invasive approach, however, remains a 2D methodology where sub-surface BSE-SEM offers insight into the first porous layer only (i.e. the porous layer at the top of the epitaxial stack, furthest from the substrate, which is the first to be porosified during the onset of the electrochemical etching process) \cite{sarkar_2024}.

This study utilises serial-section tomography in a focused ion beam scanning electron microscope (FIB-SEM) for volumetric reconstructions of three 5-pair porous GaN-on-Si DBRs, being the same materials which were first studied by Ghosh \textit{et al.} \cite{Ghosh_Sarkar_Frentrup_Kappers_Oliver_2024}. We note that details of the reflectance characteristics of these samples are discussed at length by Ghosh \textit{et al.}, to which readers who are interested in these application-relevant measurements are referred. In serial-section tomography, a dataset of aligned cross-sectional SEM images is collected by using ion beam slicing to sequentially mill the surface by a precise thickness between frames. Such tomographs, when aligned and registered, can thus provide nanoscale three-dimensional (3D) reconstructions of morphological features across microscale fields of view. This technique provides unprecedented insight into the 3D porous morphology of our DBR samples and that of porous materials generally \cite{Seo_Ha_Yoo_Kim_Lee_Kim_Kim_Cha_Kim_Jeong_et_al_2024}.

Tomographs were captured on three porous DBR samples, each cleaved from the same epitaxial wafer and porosified with the same defect-driven procedure but with different etching voltages of 5\,V, 8\,V, and 10\,V. Ghosh \textit{et al.} explored the pore morphologies of these same samples using sub-surface BSE-SEM imaging, addressing the top porous layer only \cite{Ghosh_Sarkar_Frentrup_Kappers_Oliver_2024}. Crucially, our volumetric datasets allow for the appraisal of all five porous layers from the reconstructed plan-view perspective. Likewise, these tomographs allow one to track an individual nanopipe across multiple porous layers to facilitate investigations into competing etching pathways, and for the first time, assess the validity of the kebab model in 3D.

\section{Experimental}
\label{Experimental}

\subsection{Metal-Organic Chemical Vapour Deposition}
\begin{figure}[h]
    \centering
    \includegraphics[width=1.0\linewidth]{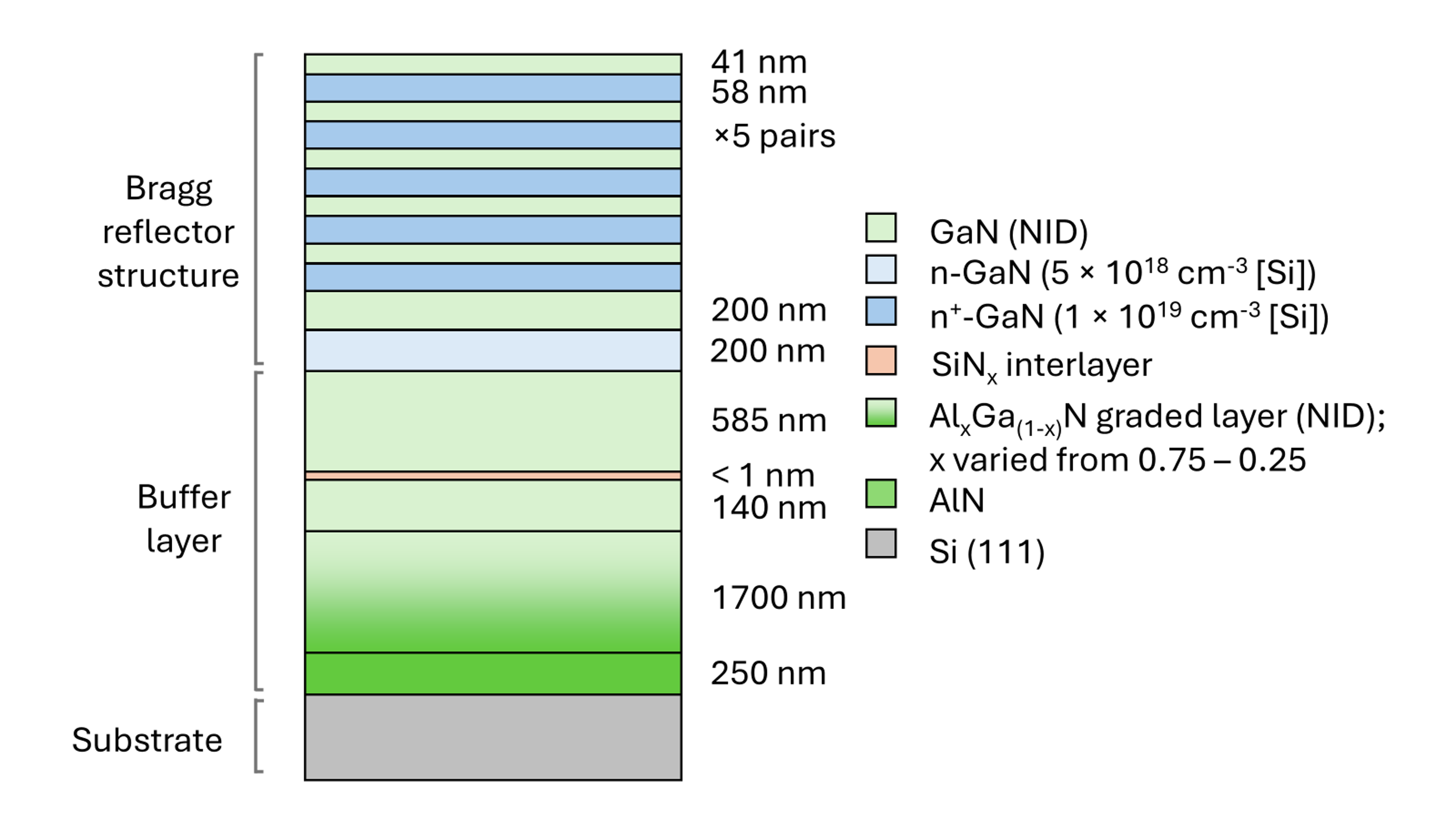}
    \caption{Simplified schematic of the as-grown epitaxial structure, prior to electrochemical porosification for the fabrication of 5-pair porous GaN-on-Si DBRs etched at 5\,V, 8\,V, and 10\,V.}
    \label{wafer schematic}
\end{figure}

All porous GaN DBRs studied in this work were prepared from the same wafer, grown by metal-organic chemical vapour deposition (MOCVD). The structure, outlined in Figure \ref{wafer schematic}, consisted of a silicon substrate, a buffer layer for lattice matching (consisting of several epilayers), and the DBR structure. The substrate was a silicon-(111) wafer with a diameter of 150\,mm and a thickness exceeding 1\,mm. The buffer layer was fabricated by depositing an initial 250\,nm AlN layer onto which a 1700\,nm layer of graded AlGaN, from Al$_{0.75}$Ga$_{0.25}$N to Al$_{0.25}$Ga$_{0.75}$N, was deposited. Onto this, a thick non-intentionally doped (NID) GaN epilayer was deposited with a thickness of 725\,nm, with a thin interlayer of silicon nitride deposited after 140\,nm of growth. Its inclusion is intended to reduce threading dislocation density by encouraging dislocation annihilation during growth and coalescence of GaN islands \cite{KAPPERS2007296}. 

Finally, the DBR structure was deposited on top of the preceding material, consisting of a Si-doped ($5 \times 10^{18} $cm$^{-3}$) n-GaN layer and NID GaN layer, each 200\,nm thick, and 5 pairs of a 58\,nm layer of highly Si-doped ($1 \times 10^{19} $cm$^{-3}$) n$^{+}$-GaN and a 41\,nm layer of NID GaN. The n$^{+}$-doped layers are selectively porosified during electrochemical etching, whilst the sub-surface n-doped layer is not porosified as part of the reflecting structure but is instead present to provide a current pathway and increase conductivity during the later stages of etching to ensure complete porosification of the deeper doped layers. This wafer was diced into smaller pieces of identical size for electrochemical etching in different conditions.

\subsection{Electrochemical Etching}
For electrochemical etching, a simple electrolytic cell was prepared, with the as-grown wafer pieces (samples) connected as the anode and a platinum disc counter electrode, with a diameter of 20\,mm cast in resin, connected as the cathode. To create the electrical contacts on the samples, a shallow scratch was first incised onto the surface at the top of each piece with a diamond-tipped scribing pen. Into this scratch, an indium metallic contact was soldered, such that direct electrical connections between the potentiostat and each of the doped layers in the sample was established. This indium contact was directly connected into the circuit, allowing electrical connection between each of the n$^{+}$-doped GaN layers for the replenishing of charge consumed during porosification and current flow in the back n-doped layer.

The DBR anode and platinum cathode were then connected to a Keithley 2400 Source Monitor controlled by a PC with Keithley Kickstart software, and both submerged into a solution of 0.25\,mol\,dm$^{-3}$ oxalic acid ($>$\,99.0\%, Sigma Aldrich). A constant voltage was then applied across the electrodes to induce electrochemical etching, such that GaN dissolved into the acid from doped layers via an electrochemical process \cite{TsengVoltage}. Active etching was evident as measured by anodic (positive) current flow; likewise, the endpoint of etching was determined after current levels had depleted to steady, negligible background levels. The three samples studied in this work were etched according to the above process at 5\,V, 8\,V, and 10\,V, respectively.

An unetched piece of the wafer was also used to assess the native threading dislocation density. This was done according to the procedure developed by Bennett \textit{et al.}, using a Bruker Dimension-Icon-Pro AFM \cite{10.1063/1.3430539}.

\subsection{Serial-section FIB-SEM Tomography}
\begin{figure}
    \centering
    \includegraphics[width=1\linewidth]{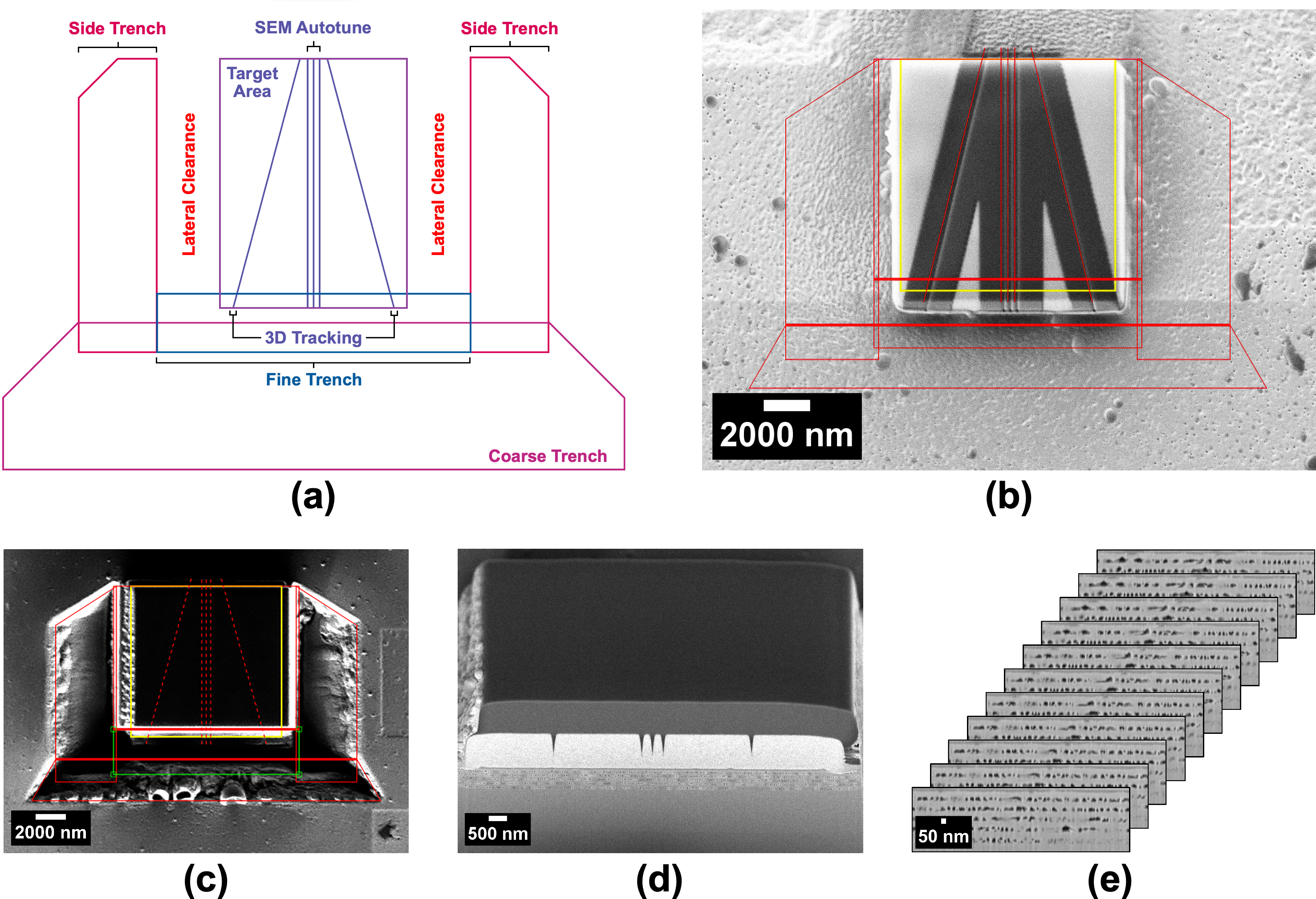}
    \caption{Outline of the serial-section FIB-SEM tomography procedure, showing a) a simplified schematic representation of the sample preparation step, b) the five lines hosted by the platinum and carbon deposited in the target area that are used for 3D tracking and SEM autotune of focus and stigmatism, c) the coarse, side, and fine trenches, milled after tracking and protection are prepared, d) cross-sectional SEM of the 5-pair porous GaN-on-Si DBR etched at 5\,V with three parallel lines in the centre as well as two converging lines on either side, hosted by the protective platinum pad and backfilled with carbon and e) 11 serial-sections from the same DBR, where the sections are related to each other with an average 5.1\,nm ion slicing thickness (Z dimension) for voxel dimensions of (2.0$\,\times$\,2.0\,$\times$\,5.1)\,nm owing to the use of 3D tracking and slice thickness interpolation.}
    \label{prep schematic}
\end{figure}

Serial-section tomography was performed using a Zeiss Crossbeam 540 FIB-SEM and the Zeiss Atlas 3D Tomography plugin for the Zeiss Atlas 5 software package. All tomographic datasets were captured according to the same conventional serial-section or `slice-and-view' tomography procedure outlined in Figure \ref{prep schematic}, differing only in the imaging conditions during acquisition and therefore final voxel dimensions.

After as-grown semiconductor wafers were porosified, cleaved samples of the three DBRs were mounted onto flat aluminium stubs using a conductive silver paste and loaded separately into the FIB-SEM instrument. For the porous GaN-on-Si DBRs etched at 5\,V and 8\,V, the regions of interest were selected arbitrarily. However, for the DBR etched at 10\,V, where it was known that pore size can far exceed the dimensions of the field of view, correlative microscopy using sub-surface BSE-SEM imaging was conducted to select a desired region of interest in advance of performing serial-section tomography such that it could be ensured that the resulting volumetric dataset encompassed one field of porosity in full. To facilitate these efforts, scratches were made on the as-etched surface using a diamond scribe, and extensive tracking images of the debris field were captured so that the region of interest imaged in the SEM could be located in the FIB-SEM afterwards.

Throughout all tomography acquisitions, the primary electron landing energy and the Ga$^{+}$ ion beam accelerating voltage were held at 2\,keV and 30\,kV, respectively, with both probe currents varied throughout. A protective 5\,\textmu m $\times$ 5\,\textmu m platinum pad was first deposited over the region of interest to reduce curtaining artefacts during serial-sectioning. This process is initially performed with electron-beam induced platinum deposition to prevent amorphisation of the as-etched porous GaN surfaces, using a probe current of 7500\,pA, producing approximately 200\,nm of platinum pad thickness. Following the initial deposition, additional platinum was overlaid using ion-beam assisted deposition with a probe current of 700\,pA, a faster process, and results in a thickness of approximately 1\,\textmu m. 

Into the platinum pad, SEM `Autotune' and `3D tracking' lines were milled with a 50\,pA ion beam. Two converging lines were milled into the platinum pad for use in 3D tracking, which are scanned in-situ to measure and sustain consistent slice thickness. Three parallel lines, running perpendicularly to the cross-sectional imaging plane, were also milled to be scanned in-situ for automated focus and stigmatism correction during acquisition, and used to align the serial-sections afterwards. All five lines are backfilled with carbon and the whole structure is further overlaid with carbon, in order to improve imaging contrast against platinum (Figure \ref{prep schematic}b). This carbon deposition is done with a 50\,pA ion beam initially, to fill the lines, followed by a thick carbon pad with a brief scan with the 1500\,nA beam. Finally, the deep trenches around the front and sides of the deposited structure were milled with a 700\,pA ion beam to ensure a clear line of sight for imaging and mitigate the effects of redeposition of milled materials onto the sample surfaces (Figure \ref{prep schematic}c). A 300\,pA ion beam was then used to finely mill the cross-sectional imaging face and ensure a planar, clean surface before tomography (Figure \ref{prep schematic}d). 

Imaging of the cross-sectional face was performed with a 65\,pA electron beam. A SE2 detector was favoured over in-column secondary electron imaging to minimise the appearance of the pore-back effect and charging artefacts. Serial-sectioning was done at an ion probe current of 100\,pA. Voxel dimensions are determined by the SEM imaging pixel size (X, Y) and the FIB slicing thickness (Z). Each slice, therefore, has a consistent X and Y dimension but a Z dimension that varies according to the width of that specific slice. Automated tracking during acquisition and post-process interpolation of slice thickness accounts for the experimental variation in slice thickness. It produces serial-section imaging data at a regular spacing to minimise distortions in tomographic reconstruction (Figure \ref{prep schematic}e). 

Continuous milling and imaging, which interlaces the electron and ion beams, promotes stability through faster serial-sectioning and the possibility of local charge neutralisation, when compared to the conventional procedure where the ion beam fully slices the cross-sectional face before the electron imaging is conducted. 

After completion of serial-section acquisition, image registration was also performed within Zeiss Atlas 5. Tomographic dataset reconstruction was conducted in Dragonfly 2024.1 (Comet Technologies Canada Inc., Montreal, Canada) \cite{Makovetsky_Piche_Marsh_2018}. 3D renders of tomographic datasets throughout are in the orthographic perspective to communicate volumetric dimensions unambiguously.

\subsection{Sub-surface BSE-SEM imaging}
Sub-surface BSE-SEM imaging of the 5-pair porous GaN-on-Si DBRs etched at 5\,V, 8\,V, and 10\,V was used for non-invasive pore morphology characterisation from the important plan-view perspective and to facilitate correlative microscopy. This imaging modality was conducted on as-etched samples without preparation according to the procedure validated by Sarkar \textit{et al.} using a Zeiss GeminiSEM 300 and the Zeiss BSD4 detector \cite{Sarkar_Adams_Dar_Penn_Ji_Gundimeda_Zhu_Liu_Hirshy_Massabuau_et_al_2024}. 

%\begin{figure}[h]
%    \centering
%    \begin{subfigure}[h]{0.243\textwidth}
%        \centering
%        \includegraphics[width=\textwidth]{Figures/Methods/3V Highlight.png}
%        \caption{3\,V}
%        \label{3V BSE}
%    \end{subfigure}
%    \hfill
%    \begin{subfigure}[h]{0.243\textwidth}
%        \centering
%        \includegraphics[width=\textwidth]{Figures/Methods/5V Highlight.png}
%        \caption{5\,V}
 %       \label{5V BSE}
 %   \end{subfigure}
 %   \hfill
%    \begin{subfigure}[h]{0.243\textwidth}
%        \centering
 %       \includegraphics[width=\textwidth]{Figures/Methods/8V Highlight.png}
 %       \caption{8\,V}
  %      \label{8V BSE}
  %  \end{subfigure}
   % \hfill
  %  \begin{subfigure}[h]{0.243\textwidth}
   %     \centering
    %    \includegraphics[width=\textwidth]{Figures/Methods/12V Highlight.png}
    %    \caption{12\,V}
    %    \label{12V BSE}
  %  \end{subfigure}
  %  \caption{Sub-surface BSE-SEM images showing the morphology of the first porous layer in 5-pair porous GaN-on-Si DBRs etched at 3\,V, 5\,V, 8\,V, and 12\,V. Micrographs are scaled to a singular scale bar. This non-invasive imaging modality generates plan-view insight into the morphology associated with the first porous layer that arises from variation in etching voltage.}
%\end{figure}

Through the use of high primary electron landing energies of 20\,keV, visualization of sub-surface pores is enabled \cite{sarkar_2024}. The image contrast generated is dominated by the morphology of the porous layer nearest to the sample surface in a DBR sample, thereby allowing non-destructive first porous layer appraisal with nanoscale spatial resolution across microscale fields of view \cite{Ghosh_Sarkar_Frentrup_Kappers_Oliver_2024}.

\newpage

\section{Results}

\begin{figure}
    \centering
    \includegraphics[width=0.9\linewidth]{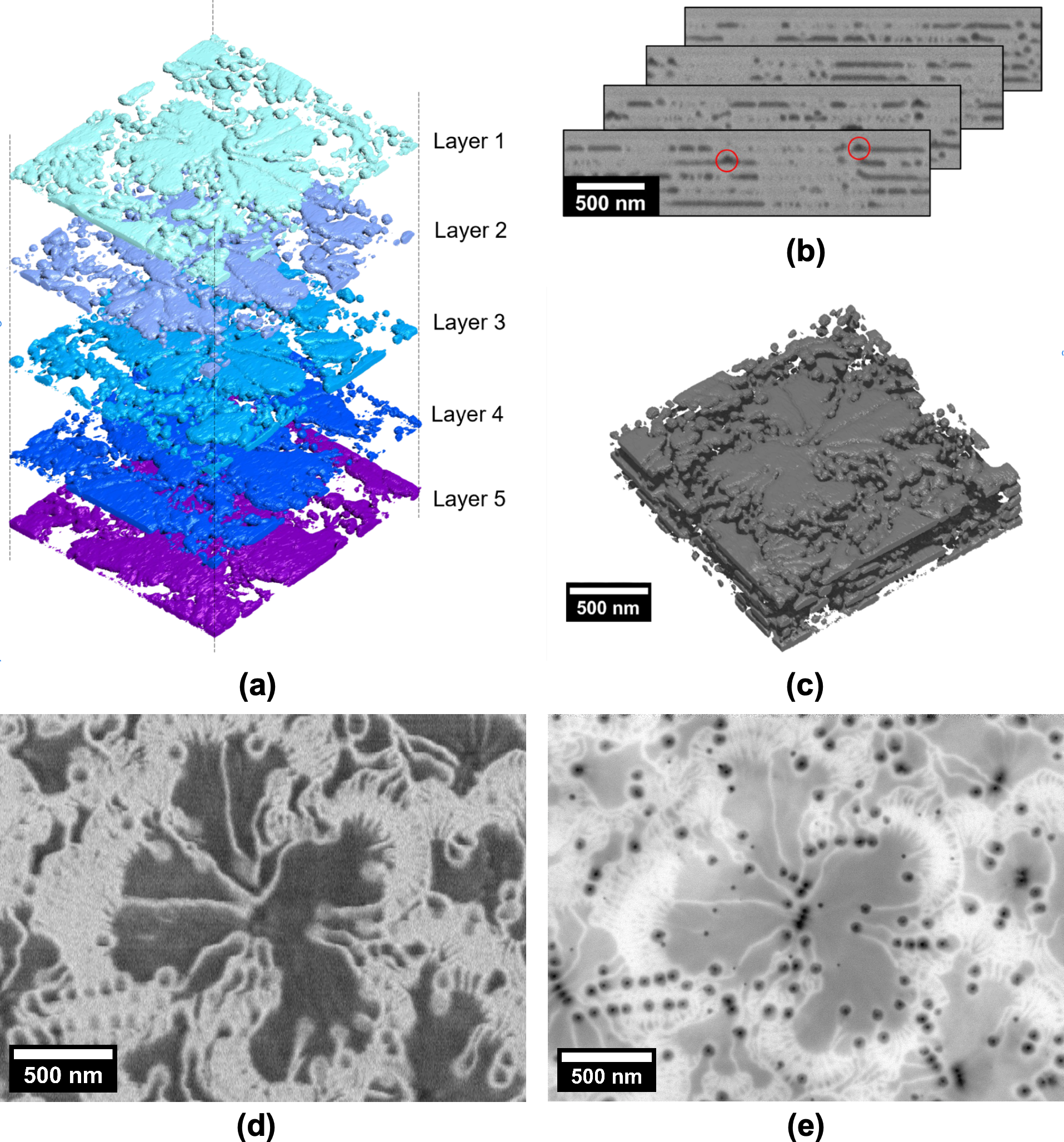}
    \caption{Serial-section FIB-SEM tomography example workflow with input serial-sections, output reconstructed plan-view images, and segmented tomograph 3D renders, shown for the 5-pair porous GaN-on-Si DBR etched at 10\,V. a)  Orthographic perspective 3D renders after tomograph segmentation with exaggerated inter-layer spacing, b) example serial-sections showing the porous region of interest, with two etched dislocation cores highlighted in red. Each of the scanning electron micrographs is captured with a 2.3\,nm $\times$ 2.3\,nm X and Y pixel size. The sections are related to each other with an 8.8\,nm Z ion slicing thickness for tomographic voxel dimensions of (2.3$\,\times$\,2.3\,$\times$\,8.8)\,nm, c) orthographic perspective 3D render of the porous layers with their real spacing, d) reconstructed plan-view of the first porous layer and e) sub-surface BSE-SEM micrograph of the first porous layer, depicting the target region of interest prior to correlative microscopy via serial-section tomography.}
    \label{early tomo workflow}
\end{figure}

The reconstruction and subsequent post-processing of datasets derived from serial-section FIB-SEM tomography experiments are summarised in Figure \ref{early tomo workflow}, all of which was performed in Dragonfly 2024.1. The input raw data from our serial-section tomography experiments are illustrated in Figure \ref{early tomo workflow}b, in the form of hundreds of cross-section SEM images, featuring a region of interest, here being the five porous/non-porous layer pairs. Image segmentation of the frames via simple intensity thresholding may be performed to extract the porous/non-porous spatial regions of the dataset and understand their structures in 3D, as illustrated by the orthographic projections in Figures \ref{early tomo workflow}a and \ref{early tomo workflow}c \cite{Otsu_1979}. After registration and post-processing of the serial-sections, virtual plan-view images can be rendered. For these, the image plane is rotated to view the structure of pores in the plane of the porous layers of the DBRs, compiling voxels obtained from real SEM frames into a reconstructed image, an example of which is given in Figure \ref{early tomo workflow}d. These plan-view reconstructions can be freely positioned, such that we can probe each porous layer and offer a more intuitive and descriptive representation of pore morphology than individual cross-sectional SEM frames.

The morphologies of pores in the three porous GaN-on-Si DBRs studied in this work can be distinguished as being in three distinct regimes, previously detailed in a study by Ghosh \textit{et al.} \cite{Ghosh_Sarkar_Frentrup_Kappers_Oliver_2024}. The spatial resolution and image contrast of tomographic datasets can thus be ascertained by comparing our reconstructed plan-view images with real plan-view images recorded using BSE-SEM using the same procedure as was employed in the Ghosh \textit{et al.} study \cite{Ghosh_Sarkar_Frentrup_Kappers_Oliver_2024}. Overall, the three tomographs were reconstructed into plan-view images that are congruous with those observed both in BSE-SEM imaging on the samples and the morphologies described by Ghosh \textit{et al.} \cite{Ghosh_Sarkar_Frentrup_Kappers_Oliver_2024}. For the tomographs of the 5\,V and 8\,V etched DBRs, we provide side-by-side comparisons of tomographic reconstructions with sub-surface BSE-SEM images of the same samples in different areas, to demonstrate the accuracy of the tomographs, shown in Supplementary Figures 1 and 2, respectively.

For the case of the 10\,V sample, directly correlative microscopy was performed, where a specific region of interest was imaged in sub-surface BSE-SEM before the serial-section FIB-SEM tomograph was captured in the same region. The results of this correlative microscopy are summarised in Figures \ref{early tomo workflow}d and \ref{early tomo workflow}e, where the same region of interest is given as a sub-surface BSE-SEM micrograph, captured before the tomography experiment, and a reconstructed plan-view image centred on the first porous layer, respectively. It is readily apparent that the tomograph captures fine details of the porous layers and that pore walls and individual nanopipes formed at threading dislocations can be correlated between the images. Detailed interpretation of the different features in comparative analysis of this type has been discussed by Sarkar \textit{et al.} for a similar porous GaN DBR grown on a sapphire substrate \cite{Sarkar_Adams_Dar_Penn_Ji_Gundimeda_Zhu_Liu_Hirshy_Massabuau_et_al_2024}.

With these tomographs, pore morphology, as well as the nanopipes formed at etched threading dislocations and associated porous fields, can readily be examined. Dislocations that etch through NID GaN layers and form porous fields in deeper, highly Si-doped GaN layers of the DBR multilayer stack are demonstrated to partially etch the bottom of the NID layer, meaning the resulting pore contains a vertical protrusion extending upwards from the doped layer. Examples of this effect, as viewed in cross-section, are highlighted by red rings in Figure \ref{early tomo workflow}b, where we see examples of triangular pores that extend into the non-porous layers. This effect can be exploited in the context of tomography by creating plan-view reconstructions not positioned \textit{inside} porous layers, but slightly above them, such that the full expansion of pores is not captured, but the porous protrusions, centred on threading dislocation cores, are captured. These `etching onset' reconstructions are invaluable, allowing for correlation of porous fields in the doped layers immediately below to individual dislocations, but also themselves serving essentially as maps of the positions of dislocations which have opened up into nanopipes to form an etching pathway \cite{sarkar_2024}. 

For these reasons, the presentation of tomographic datasets in this work will be two-fold. Presentation of each of the three datasets will consist of annotated reconstructed plan-view images, positioned both in the five porous layers and in their corresponding regions on non-porous layers (`onset frames') positioned slightly above them (figures \ref{5VFIBSEM}, \ref{8VFIBSEM} and \ref{10VFIBSEM}), and also in the form of animated fly-through videos, showing both cross-sectional and reconstructed plan-view perspectives (Videos 1, 2 and 3). This allows for datasets to be provided in full and also for important features of discussion to be highlighted through annotations. We note that for each video, the cross-sectional frames are animated in the order in which they were captured (from the front of the region of interest to the back) and the reconstructed plan-view frames are animated from the top of the region of interest vertically downward (i.e. the first porous layer appears chronologically first). The 5\,V and 8\,V tomographs are also presented as orthographic projections of segmented pore structures, similarly to Figure \ref{early tomo workflow}c featuring the 10\,V sample tomograph, in Supplementary Figure 3.

\subsection{5-pair porous GaN-on-Si DBR etched at 5\,V}

(The authors request that Video 1, titled `Video 1 - 5-pair porous GaN-on-Si DBR etched at 5V.mp4', is embedded at this position in the text with the associated caption `Video 1 - Tomographic dataset captured on the DBR sample etched at 5\,V, shown in full as the raw cross-sectional dataset followed by the same raw data viewed from the plan-view reconstructed perspective')

\begin{figure}
    \centering
    \includegraphics[width=1\linewidth]{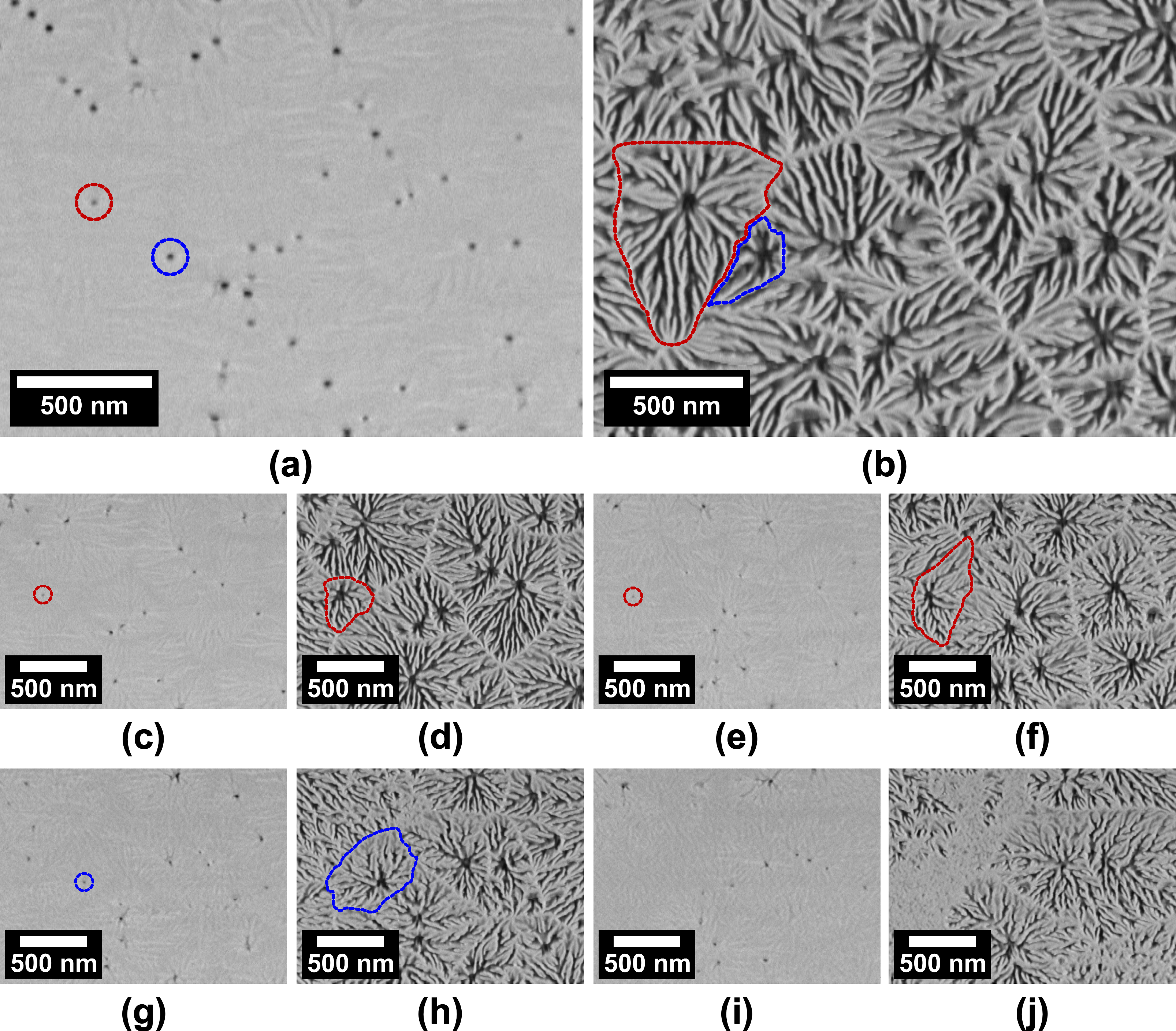}
    \caption{Reconstructed plan-view images as a function of depth, derived from serial-section FIB-SEM tomography of the 5-pair porous GaN-on-Si DBR etched at 5\,V. Images are extracted positioned at a) onset of layer 1, b) layer 1, c) onset of layer 2, d) layer 2, e) onset of layer 3, f) layer 3, g) onset of layer 4, h) layer 4, i) onset of layer 5 and j) layer 5. `Dislocation 1' is highlighted in red throughout, alongside `dislocation 2' in blue.}
    \label{5VFIBSEM}
\end{figure}
 
Video 1 shows the full tomographic dataset, captured on the sample etched at 5\,V, as both raw input frames and associated horizontal plan-view images. For this tomograph, the voxel dimensions (with average Z thickness) are (2$\,\times$\,2\,$\times$\,5.1)\,nm. Inferring the pore morphology using the cross-sectional frames is difficult, but the horizontal plan-view images show the in-plane pore morphology clearly. Annotated plan-view reconstructions of the tomograph of the sample etched at 5\,V are presented in Figure \ref{5VFIBSEM}. Here, each of the five porous layers is depicted with an associated image that shows the onset of electrochemical etching for that highly Si-doped GaN layer. For each porous layer, there is a variation in the observed image contrast that can be attributed to pore morphology. Regions where the branched pores appear darker are related to the depth within the reconstructed plan-view image, with the back of the pore being further from the reconstructed oblique \cite{Reimers_Safonov_Yakimchuk_2019}.

The reconstructions at the centres of the five porous layers, being Figures \ref{5VFIBSEM}b, \ref{5VFIBSEM}d, \ref{5VFIBSEM}f, \ref{5VFIBSEM}h, and \ref{5VFIBSEM}j, respectively, demonstrate the expected morphologies observed in sub-surface BSE-SEM imaging (Supplementary Figure 1). The in-plane morphologies consist of central porous spots where threading dislocations have originally etched into doped layers, and long fine-tipped branches emanating from these outward into the layer in lateral directions. Porous fields from different dislocations do not intersect or coalesce - the pores branching from dislocation spots therefore form distinct `cells', where singular branched regions are separated by pore walls of solid GaN. These porous cells are demonstrated to continue throughout the stack, with all five porous layers featuring this morphology. 

Notably, the fifth porous layer, shown in Figure \ref{5VFIBSEM}j, has a sharp drop-off in porosity, an effect observed previously via cross-sectional SEM in deeper porous layers of 5-pair porous GaN-on-Si DBRs etched at low voltages \cite{Ghosh_Sarkar_Frentrup_Kappers_Oliver_2024}. Here, we now understand that this reduction in porosity is brought about by a smaller number of porous cells forming, with the ones that do form unable to expand to fill space and fully porosify the layer. 

Onset etching frames for the five porous layers are given in Figures \ref{5VFIBSEM}a, \ref{5VFIBSEM}c, \ref{5VFIBSEM}e, \ref{5VFIBSEM}g, and \ref{5VFIBSEM}i, respectively, extracted from a position slightly above the reconstructions of their corresponding porous layers. At 5\,V, the lowest etching voltage studied, the onset frames consist of relatively small black spots which expand outward into the cells when moving down through the stack, as is clear in Video 1.

Two etched threading dislocations in the field of view of the tomograph are highlighted in red and blue throughout Figure \ref{5VFIBSEM}. These dislocations are black spots in the onset layers and form branched cells in porous layers, centred on the same position as the previous dark spot. For the first porous layer, as highlighted in Figure \ref{5VFIBSEM}a and \ref{5VFIBSEM}b, both dislocations obey this behaviour. In the second porous layer, however, we observe that dislocation 1 (highlighted in red throughout) can be identified as both a dark spot in Figure \ref{5VFIBSEM}c and a cell centre in Figure \ref{5VFIBSEM}d, but that dislocation 2 (highlighted in blue throughout) is no longer present in either. Instead, in the porous layer we find that the region of space previously occupied by the pore from dislocation 2 has now been etched by a porous field emanating from a different neighbouring dislocation, implying that the nanopipe forming on dislocation 2 has been `undercut'. It can be concluded from this that dislocation 1 has been etched into a nanopipe and formed a porous layer in the first two doped layers according to the conventional expected `kebab' behaviour, but that dislocation 2 deviates from this by `switching off' after forming a pore in the first layer. 

Examination of the third porous layer in Figures \ref{5VFIBSEM}e and \ref{5VFIBSEM}f demonstrates the same behaviour as the previous layer; dislocation 1 continues to etch, forming both a spot on the onset layer and a porous cell in the doped layer (highlighted), whilst dislocation 2 remains absent. In the fourth porous layer, however, the behaviour changes yet again, and in Figures \ref{5VFIBSEM}g and \ref{5VFIBSEM}h we observe that dislocation 1 is now absent whilst dislocation 2 has reappeared, resuming its behaviour as an etching pathway forming a nanopipe and associated porous field (highlighted). Furthermore, in the fifth porous layer outlined in Figures \ref{5VFIBSEM}i and \ref{5VFIBSEM}j, neither dislocation 1 nor 2 is present, meaning both are `inactive' as etching pathways through the fifth layer. 

Thus, dislocation 1 is demonstrated to act as an etchant pathway throughout the first three porous layers, before ceasing to do so in the fourth and fifth, whilst dislocation 2 has been etched in the first and fourth porous layers only. Both dislocations exhibit distinct behaviour relative to one another and also deviate from the conventional `kebab' behaviour described previously.

\subsection{5-pair porous GaN-on-Si DBR etched at 8\,V}

(The authors request that Video 2, titled `Video 2 - 5-pair porous GaN-on-Si DBR etched at 8V.mp4', is embedded at this position in the text with the associated caption `Video 2 - Tomographic dataset captured on the DBR sample etched at 8\,V, shown in full as the raw cross-sectional dataset followed by the same raw data viewed from the plan-view reconstructed perspective')

\begin{figure}
    \centering
    \includegraphics[width=1\linewidth]{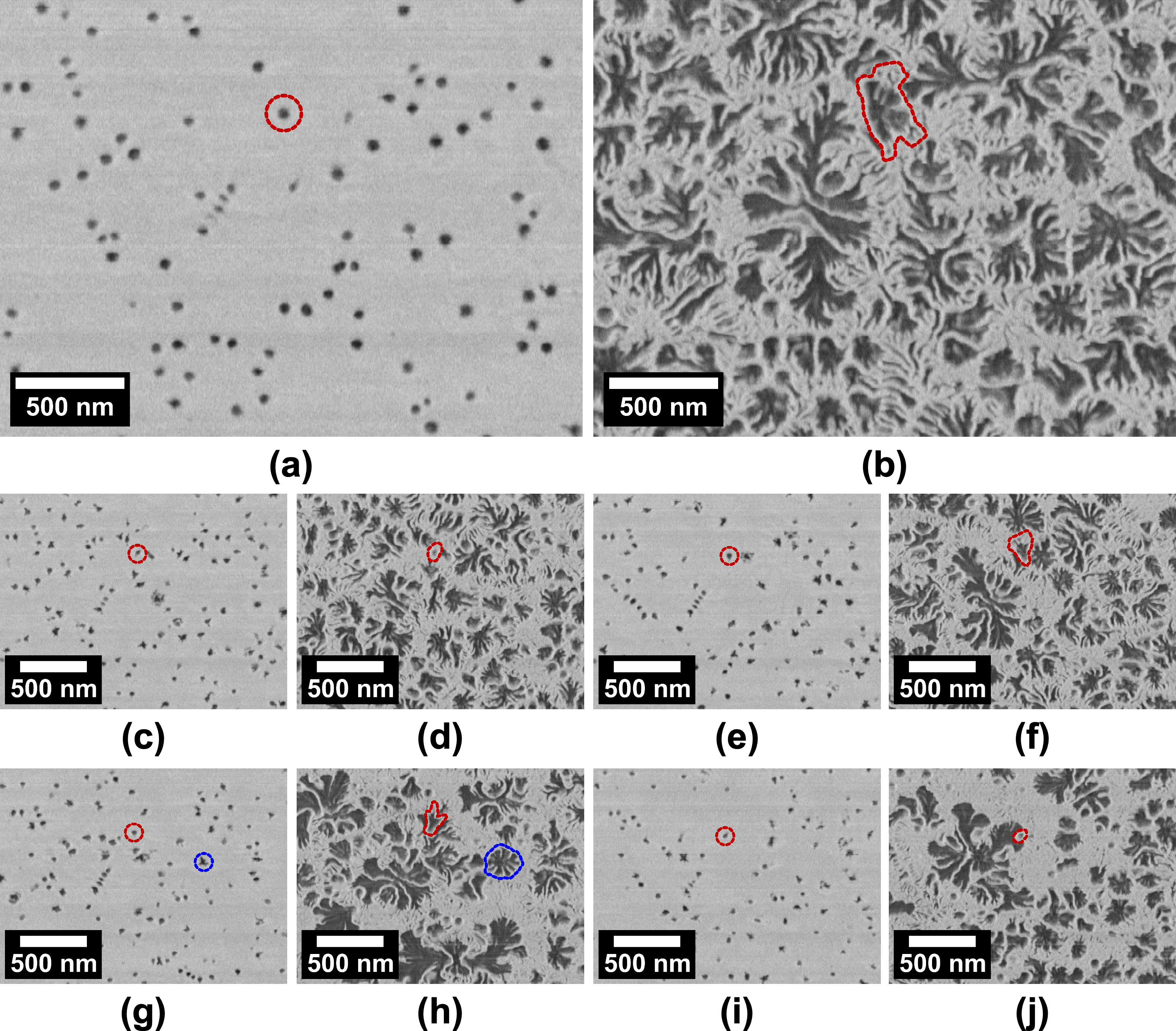}
    \caption{Reconstructed plan-view images as a function of depth, derived from serial-section FIB-SEM tomography of the 5-pair porous GaN-on-Si DBR etched at 8\,V. Images are extracted positioned at a) onset of layer 1, b) layer 1, c) onset of layer 2, d) layer 2, e) onset of layer 3, f) layer 3, g) onset of layer 4, h) layer 4, i) onset of layer 5 and j) layer 5. `Dislocation 3' is highlighted in red throughout, alongside `dislocation 4' in blue.}
    \label{8VFIBSEM}
\end{figure}

The tomographic dataset for the sample etched at 8\,V is presented in full in Video 2 in the same format as Video 1, and again as a series of annotated plan-view images in Figure \ref{8VFIBSEM}. For this tomograph, the voxel dimensions (with average Z thickness) are (2.3$\,\times$\,2.3\,$\times$\,5.7)\,nm. It is readily apparent that the pore morphologies in porous layers, given in Figures \ref{8VFIBSEM}b, \ref{8VFIBSEM}d, \ref{8VFIBSEM}f, \ref{8VFIBSEM}h, and \ref{8VFIBSEM}j, are different to the previous sample etched at 5\,V. Here, porous cells have wider arms and do not display the fine-tipped branches, but instead feature broader, more cavernous porous fields. A comparison of the pore morphology for this sample as captured by the tomograph and also by sub-surface BSE-SEM imaging is given in Supplementary Figure 2, where once again the morphologies demonstrated by the two techniques are commensurate, indicating that the tomograph has accurately reconstructed the structure.

Onset frames, given in Figures \ref{8VFIBSEM}a, \ref{8VFIBSEM}c, \ref{8VFIBSEM}e, \ref{8VFIBSEM}g, and \ref{8VFIBSEM}i, have been extracted from a greater height above their porous layers than in the previous sample, since the extent of nanopipe expansion is greater owing to the elevated etching voltage. They appear as dark spots which are larger than in the previous dataset. There is also a striking increase in the density of threading dislocations, which have etched in each of the five porous layers compared to the 5\,V sample given in Figure \ref{8VFIBSEM}, clearly visible from the larger number of dark spots present in the field of view on the onset frames. 

Two additional dislocations observed in Figure \ref{8VFIBSEM} have been highlighted in red and blue, each with unique etching behaviour. The first, denoted `dislocation 3' (highlighted in red), appears as an etching pathway and resulting pore in all five porous layers and their respective onset frames. We note that the pores in formed in porous layers by this dislocation are smaller, due to a large diversity of pore sizes across this sample. This dislocation hence follows conventional kebab behaviour, forming a central nanopipe through the entire stack and etching in all five porous layers. The second dislocation, denoted `dislocation 4' (highlighted in blue), appears only in the fourth porous layer/onset frame, where it forms a porous field (Figures \ref{8VFIBSEM}g and Figure \ref{8VFIBSEM}h). In all other layers, there is no dislocation present as a black spot in the onset frame or porous field centred on the corresponding region. Hence, this dislocation only etches in one layer of the stack, being the fourth layer. The two dislocations highlighted in this tomograph show more distinct behaviour again, with an instance of a dislocation that has etched to form a conventional structure predicted by the kebab model (dislocation 3) and a dislocation that has not etched on the surface (not forming a pore in the first layer) but activates in the fourth layer (dislocation 4).

\subsection{5-pair porous GaN-on-Si DBR etched at 10\,V}

(The authors request that Video 3, titled `Video 3 - 5-pair porous GaN-on-Si DBR etched at 10V.mp4', is embedded at this position in the text with the associated caption `Video 3 - Tomographic dataset captured on the DBR sample etched at 10\,V, shown in full as the raw cross-sectional dataset followed by the same raw data viewed from the plan-view reconstructed perspective')

\begin{figure}
    \centering
    \includegraphics[width=1\linewidth]{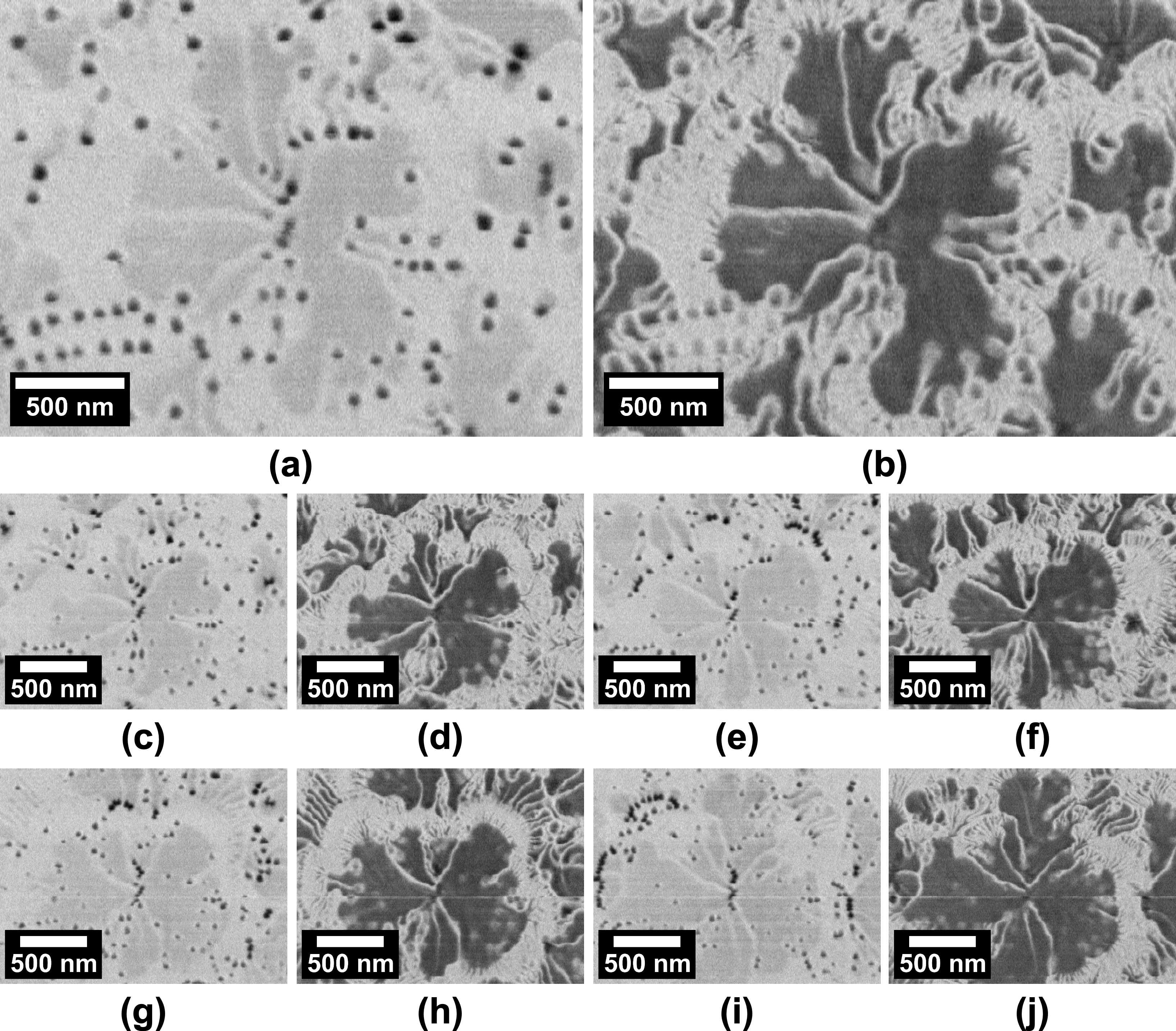}
    \caption{Reconstructed plan-view images as a function of depth, derived from serial-section FIB-SEM tomography of the 5-pair porous GaN-on-Si DBR etched at 10\,V. Images are extracted positioned at a) onset of layer 1, b) layer 1, c) onset of layer 2, d) layer 2, e) onset of layer 3, f) layer 3, g) onset of layer 4, h) layer 4, i) onset of layer 5 and j) layer 5.}
    \label{10VFIBSEM}
\end{figure}

Finally, the tomograph of the DBR etched at 10\,V is presented in Video 3 and Figure \ref{10VFIBSEM} in the same configuration of animated video and annotated plan-view reconstructions employed previously. For this tomograph, the voxel dimensions (with average Z thickness) are (2.3$\,\times$\,2.3\,$\times$\,8.8)\,nm. The plan-view reconstruction of the first porous layer, featured previously as Figure \ref{early tomo workflow}d, is reproduced in Figure \ref{10VFIBSEM}b, now in the context of its onset and additional deeper porous layers. As mentioned previously, the tomograph features the expected `lily-pad' morphology from our correlative microscopy and prior work, despite the coarser voxel dimensions compared to the other tomographs \cite{Ghosh_Sarkar_Frentrup_Kappers_Oliver_2024,Sarkar_Adams_Dar_Penn_Ji_Gundimeda_Zhu_Liu_Hirshy_Massabuau_et_al_2024}.

Moving down through the individual porous layers in Figures \ref{10VFIBSEM}b, \ref{10VFIBSEM}d, \ref{10VFIBSEM}f, \ref{10VFIBSEM}h, and \ref{10VFIBSEM}j, it is apparent that the large pore at the centre of the first porous layer in Figure \ref{10VFIBSEM}b dominates and persists through the deeper layers, with each layer featuring a large central pore that, despite changing shape and moving, forms a vertical stack of large pores. This stack, at a glance, appears to follow kebab-like behaviour, where we have discovered five overlaying pores in the highly Si-doped layers that emanate from a common centre. However, through examination of the onset frames in Figures \ref{10VFIBSEM}a, \ref{10VFIBSEM}c, \ref{10VFIBSEM}e, \ref{10VFIBSEM}g, and \ref{10VFIBSEM}i, it becomes clear that this large porous field is actually comprised of porous fields emanating from several dislocations, rather than the single dislocation highlighted in conventional kebab behaviour. These dislocations in fact have their own separate activation/deactivation behaviour.

It is apparent from the inspection of the three tomographs that the kebab model for defect-driven DBR porosification offers an over-simplified view, as instances that both seemingly corroborate it and contradict it are demonstrated. To formulate a more accurate and sophisticated model to account for the behaviour of dislocations as etching pathways, it is helpful to analyse which dislocations are active as etching pathways, and in which layers this is true, but do so en masse across the numerous dislocations present in all three tomographs rather than through specific instances associated with individual dislocations. 

\section{Discussion}
\begin{figure}[h]
    \centering
    \includegraphics[width=1\textwidth]{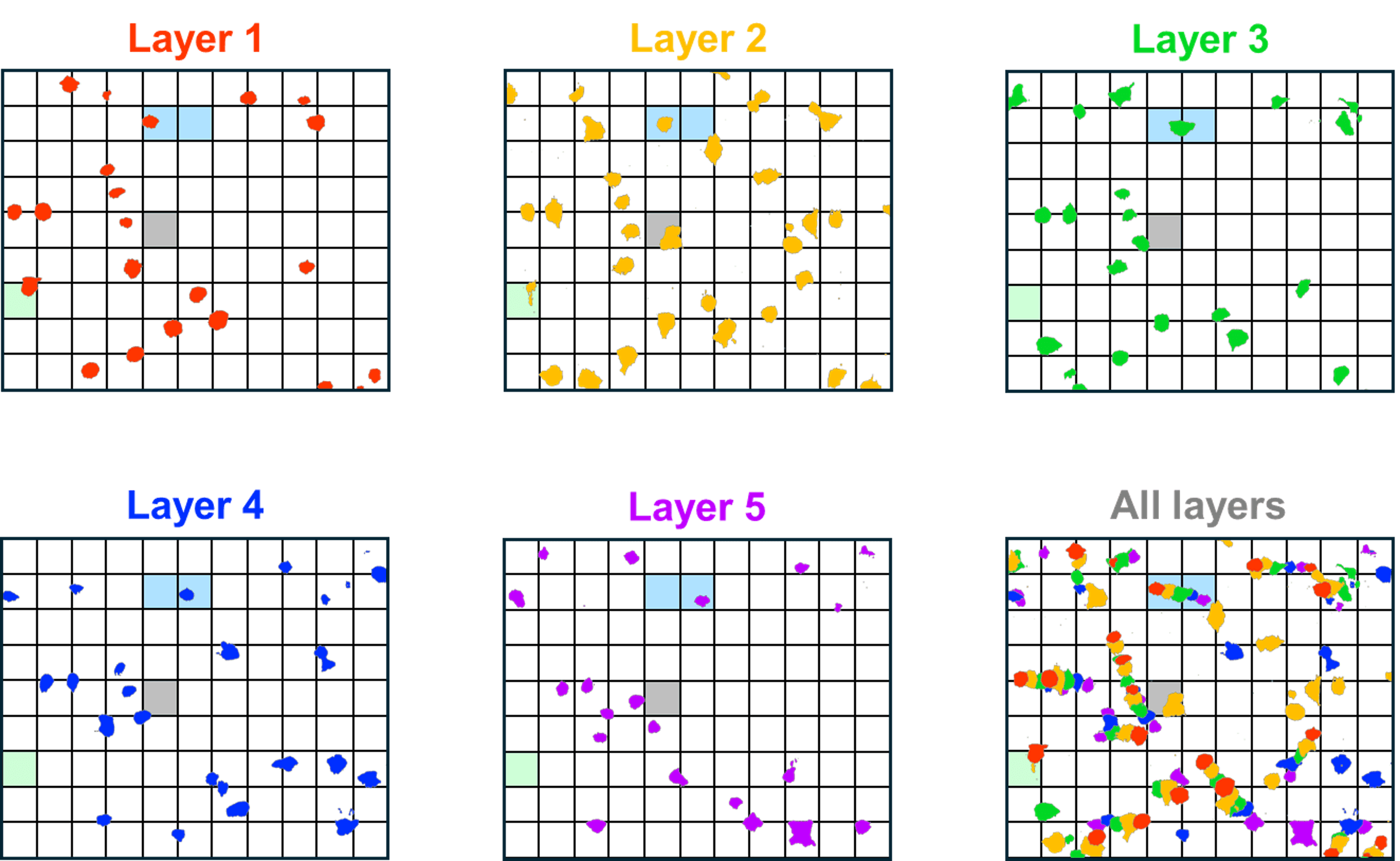}
    \caption{A series of images cropped and recoloured from the plan-view reconstructions of the 5-pair porous GaN-on-Si DBR etched at 8\,V first shown in Figure \ref{8VFIBSEM} and derived from serial-section FIB-SEM tomography. All grid line separations are 100\,nm. The onset of five porous layers is individually shown, with an additional sixth image that overlays all layers. Highlighted squares show regions containing threading dislocations with different activation behaviours during defect-driven electrochemical etching.}
    \label{confetti}
\end{figure}

In this section, we demonstrate a novel analysis method to assess threading dislocations as etching pathways at large using the onset frames of the reconstructed tomographs. Each onset frame, when segmented, recoloured and overlaid to the other four onset frames in the tomograph, can show which dislocations are active in each porous layer. Overlaying the maps allows for individual threading dislocations to be identified and tracked, with each of the layers it etched in being identified as the coloured maps in which it appears. An example demonstrating this procedure is given in Figure \ref{confetti}, showing the same cropped region of the five onset frames of the 8\,V tomograph (given previously in Figure \ref{8VFIBSEM}), as well as the appearance of the segmented and recoloured frames when all are overlaid. 

Threading dislocations in GaN-on-Si epitaxy have been demonstrated to incline to angles off from the vertical c-axis direction \cite{Inotsume_Kokubo_Yamada_Onda_Kojima_Ohara_Harada_Tagawa_Ujihara_2020}. This threading dislocation inclination may be associated with tensile strain induced by silicon dopant atoms and/or lattice mismatch strain \cite{Inotsume_Kokubo_Yamada_Onda_Kojima_Ohara_Harada_Tagawa_Ujihara_2020,inclineddislocations}. The inclination of individual dislocation cores away from the vertical c-axis is also clearly demonstrated in our tomographic datasets. In the context of Figure \ref{confetti}, this inclination presents as lateral displacement between the horizontal plan-view images. The light blue highlighted squares feature one dislocation core which, due to its inclination angle, translates to the right in the field of view when moving down the porous layers, and appears as a smear in the overlaid figure. Since the core is visible in all five layers with a consistent displacement, it demonstrates that this dislocation is active in all five porous layers and forms a nanopipe running through the entire DBR structure. The pale green highlighted space shows a dislocation core, which is active in the first and second layers and not active in the remaining ones, from which no pores emanate in the final three layers. This dislocation has been activated on the surface and deactivated at the bottom of the second porous layer. Finally, the grey highlighted space shows a dislocation core which etched only in the second layer, not being used as a pathway for the first, third, fourth and fifth layers.

According to the kebab model, we would expect the threading dislocations to all act as etching pathways in every porous layer, which would present here as each dislocation present as a 5-layer coloured overlay (either overlapping for a non-inclined dislocation or smeared for an inclined dislocation). Clearly, as seen in the light blue squares of Figure \ref{confetti}, there are dislocations that appear like this and therefore there are dislocations that form conventional `kebab' pore structures. However, this is not the case for all the dislocations present in Figure \ref{confetti} or in the tomographs as a whole. Using this methodology for the full field of view of the three tomographs, each dislocation that has acted as an etching pathway (and is therefore active) in any of the layers can be tracked, and the specific layers for which it is active can be identified. This process then generates datasets for each tomograph of every dislocation, and which layers they are active within.

\subsection{Cascade model}
Before presenting and discussing the statistical datasets for threading dislocation etching, based on analyses like those shown above, it is essential to provide a new mechanistic evaluation of the defect-driven etching process and explain how the etching pathways and associated pore morphology observed through serial-section FIB-SEM tomography can physically arise. 

It has been demonstrated in Figure \ref{confetti}, in the light blue squares, that there are etched threading dislocations which form conventional `kebabs', with a central etched nanopipe and five porous fields in highly Si-doped layers. Figure \ref{confetti} also demonstrates a second etched dislocation, in the pale green squares, which is active in the first and second layers, then inactive beyond this point. Such a dislocation, which does not follow conventional kebab behaviour, can still be described as having a `continuous' etching sequence despite the observation of premature nanopipe termination, as the layers where it acts as an active etchant pathway (layers one and two) are all adjacent. 

However, in Figure \ref{5VFIBSEM}, dislocation 2 serves as an active etching pathway in the first layer, remains inactive in the second and third layer, and reactivates in the fourth layer. Such a dislocation may be distinguished from those previously described as having an `interrupted' etching sequence, since the layers for which it is an active etching pathway (being layers one and four) are not adjacent, but separated by a layer in which the dislocation is inactive.  In addition to these discontinuities in nanopipe formation, and in contrast to the previous two etching pathways, this pathway initially activates sub-surface in this otherwise surface-initiated electrochemical etching process, further deviating from the pore morphology as described by the simple kebab model for defect-driven etching. Indeed, assessing the active layers of dislocations in all three tomographs using the methodology outlined above, a diverse range of activation/deactivation behaviour can be observed. 

\begin{figure}
    \centering
    \includegraphics[width=1\linewidth]{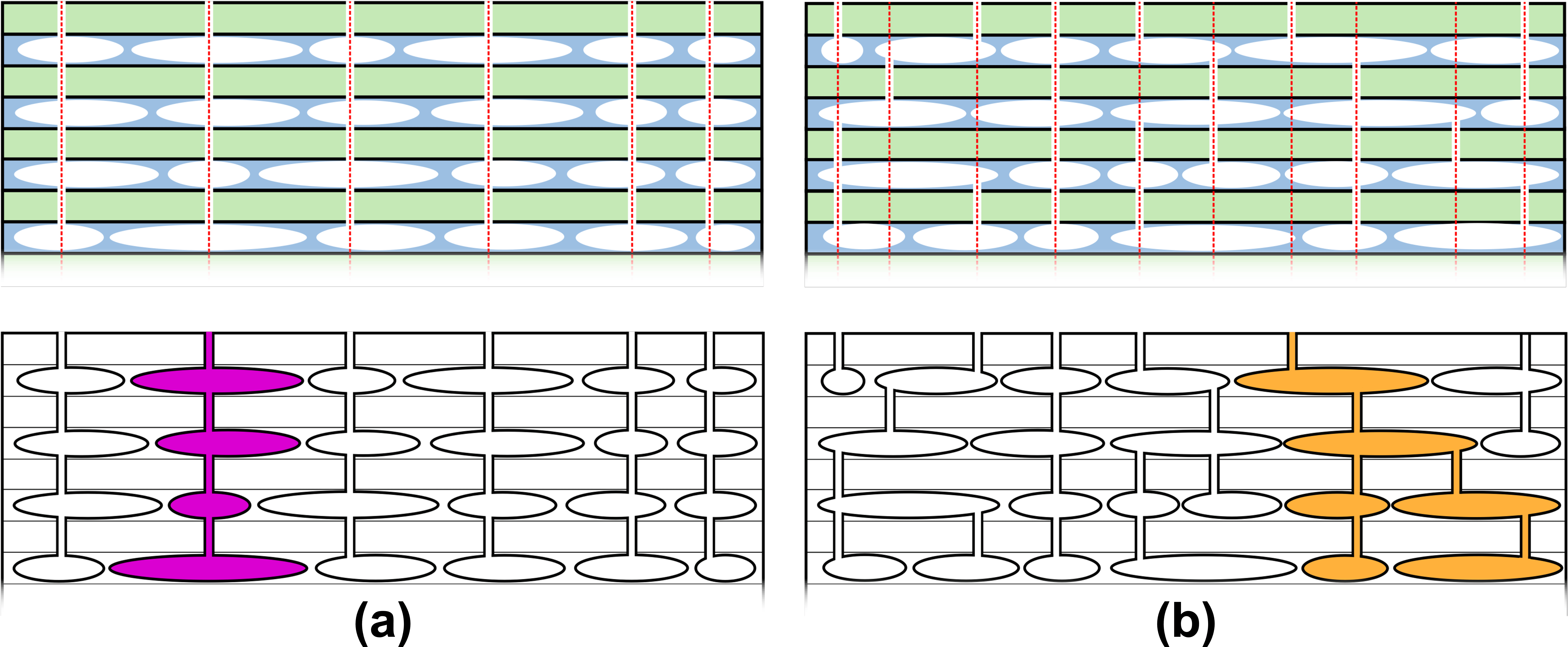}
    \caption{A schematic diagram of a) the conventional `kebab model', as contrasted to b) the improved `cascade model', for the proliferation of defect-driven etching. Red lines represent threading dislocation cores, and blue and green layers represent highly Si-doped layer and NID layers, respectively. Pore structures are highlighted below the schematics, with coloured structures demonstrating the differences between the models.}
    \label{cascade schem}
\end{figure}

We thus propose a `cascade' model of defect-driven etching of porous GaN DBRs. In the kebab model, each dislocation core etches to form an isolated and distinct porous structure, consisting of a central nanopipe with emanating pores in highly Si-doped layers. In our cascade model, we propose that several distinct dislocation cores etch into nanopipes through different layers in the stack to form one combined network of interacting porous domains. A schematic diagram comparing the existing and proposed models is given in Figure \ref{cascade schem}, showing four porous (blue) and non-porous (green) layers with pores etching via dislocation cores (red). 

%\begin{figure}[h]
%    \centering
%    \begin{subfigure}[h]{0.495\textwidth}
%        \centering
%        \includegraphics[width=\textwidth]{Figures/Discussion/Kebab model schematic.png}
%        \caption{The `kebab' model}
%        \label{KSCHEM}
%    \end{subfigure}
%    \hfill
%    \begin{subfigure}[h]{0.495\textwidth}
%        \centering
 %       \includegraphics[width=\textwidth]{Figures/Discussion/Cascade model schematic.png}
 %       \caption{The `cascade' model}
 %       \label{CSCHEM}
 %   \end{subfigure}    
 %   \caption{A schematic diagram of the improved `cascade model', as contrasted to the conventional `kebab model', for the proliferation of defect-driven etching. Red lines represent threading dislocation cores, and blue and green layers represent highly Si-doped layer and NID layers, respectively. Pore structures are highlighted below the schematics, with coloured structures demonstrating the differences between the models.}
 %   \label{cascade schem}
%\end{figure}

Figure \ref{cascade schem}b shows the larger porous structures formed by etched dislocations, which themselves do not etch through all layers, in contrast with the conventional kebab model. A dislocation core that does not activate at the surface and hence has no associated porosity in the first porous layer, may be activated if a nearby dislocation core that is active in the first layer produces a pore that overlaps with that dislocation. This behaviour is shown in Figure \ref{cascade schem}b as etched dislocation cores that do not form nanopipes on the surface, but do so after a pore emanating from a nearby etched dislocation intersects the inactive dislocation. Etching can then proceed down either of the available dislocations through the NID layer beneath, allowing dislocations that were previously inactive to become the new dominant pathway. Likewise, a dislocation can be deactivated in a similar manner, if the etching of a nearby dislocation produces a porous region in a doped layer before that dislocation has etched fully through the NID layer above. In this way, the undercutting of a dislocation by another one nearby that reaches the porous layer earlier, allows for a depletion region \cite{griffin_patel_zhu_langford_kamboj_ritchie_oliver_2020} to block the undercut dislocation from proceeding any further. Rather than forming distinct porous structures, therefore, threading dislocations act as etching pathways which overlap and interact, forming complex networks of interconnected dislocations and porous fields in doped layers. 

The cascade model, as described here, is the only explanation for the presence of etched dislocations which can either follow conventional kebab behaviour, form otherwise continuous structures or form distinct interrupted structures. To reiterate, the cascade model can also encompass conventional kebab behaviour, with 5-layer porous structures being valid and experimentally observed etching pathways. Figure \ref{cascade schem} includes such a structure on the fourth dislocation from the left. The cascade model is hence an enhancement of the kebab model, rather than a complete framework replacement.

\subsection{Dislocation statistics}

Returning to the statistics of threading dislocations as etching pathways with a mechanistic evaluation of how such structures may form, we can extract several parameters to compare behaviour across the etching voltages. First, neglecting briefly the specific porous layers for which dislocations are active (etched), focusing only on the total number of layers for which they are active, an arithmetic mean for the number of porous layers in which dislocations are active for the three samples can be found. This average number of doped layers for which dislocations are active in the 5-pair porous GaN-on-Si DBRs etched at 5\,V, 8\,V, and 10\,V is 1.75, 2.78, and 2.82, respectively. Furthermore, the total density of active dislocations, neglecting which layers or how many layers they are active in, can be given for each sample. For the 5\,V, 8\,V, and 10\,V porous DBRs, the total active dislocation density is $1.9\times 10 ^{9}$\,cm$^{-2}$, $3.5\times 10 ^{9}$\,cm$^{-2}$, and $4.4\times 10 ^{9}$\,cm$^{-2}$, respectively. These two increasing trends both relate to the greater porosity observed in porous layers of DBRs etched at higher etching voltages, with the latter (the total density of participating active dislocations) illustrating that a given dislocation in the sample has a greater probability of etching at higher etching voltage, and the former (the average number of doped layers for which a dislocation is active) showing that the number of layers in which those dislocations are active also increase with increasing etching voltage. Both of these can be associated with the increased etching voltage leading to faster and more extensive porosification via the hole-mediated process of highly Si-doped GaN electrochemically etching in oxalic acid, a well-documented feature of porosification of GaN in both DBR structures \cite{Ghosh_Sarkar_Frentrup_Kappers_Oliver_2024} and monolithic doped layers \cite{TsengVoltage}. 

\begin{figure}
    \centering
    \includegraphics[width=1\linewidth]{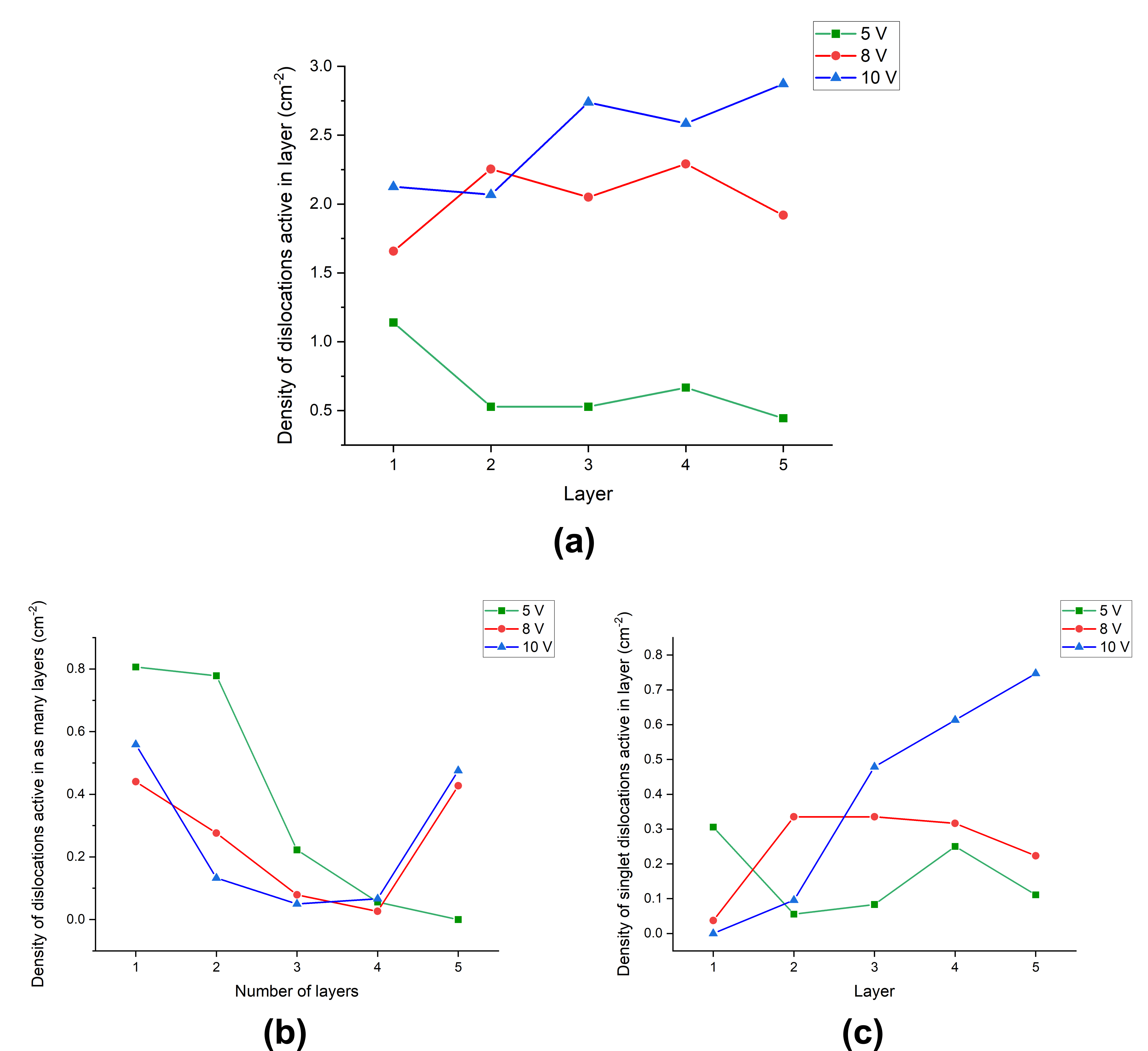}
    \caption{Dislocation statistics for 5-pair porous GaN-on-Si DBRs etched at 5\,V, 8\,V, and 10\,V, given as a) Active dislocation density for different layers in samples at different etching voltages, b) Active dislocation density for different numbers of active layers at different etching voltages and c) Density of dislocations that have etched in only one layer ('singlets') at different etching voltages.}
    \label{graphs}
\end{figure}

We can individually appraise the active dislocation density (`ADD') for each porous layer rather than across all five porous layers. This local value can then be compared across the different layers and across the different etching voltages, as given in Figure \ref{graphs}a. Likewise, dislocation statistics may be expressed in terms of the number of layers in which dislocations are active; the density of dislocations active in a specific number of layers, disregarding the particular layers they are in, is given in Figure \ref{graphs}b.

%\begin{figure}[h!]
%    \centering
%    \begin{subfigure}[h]{0.6\textwidth}
%        \centering
%        \includegraphics[width=\textwidth]{Figures/Discussion/ADD per layer plot.png}
%        \caption{ADD by layers}
%        \label{7a}
%    \end{subfigure}
 %   \par\smallskip
 %   \begin{subfigure}[h]{0.495\textwidth}
 %       \centering
 %       \includegraphics[width=\textwidth]{Figures/Discussion/ADD per number of layers plot.png}
 %       \caption{ADD by number of layers}
 %       \label{7b}
 %   \end{subfigure}
 %   \hfill
 %   \begin{subfigure}[h]{0.495\textwidth}
 %       \centering
 %       \includegraphics[width=\textwidth]{Figures/Discussion/Singlets plot.png}
  %      \caption{Singlet distributions}
  %      \label{7c}
  %  \end{subfigure}    
  %  \caption{Dislocation statistics for 5-pair porous GaN-on-Si DBRs etched at 5\,V, 8\,V, and 10\,V, given as a) Active dislocation density for different layers in samples at different etching voltages, b) Active dislocation density for different numbers of active layers at different etching voltages and c) Density of dislocations that have etched in only one layer ('singlets') at different etching voltages.}
  %  \label{graphs}
%\end{figure} 

Ghosh \textit{et al.} reported an increasing trend in the number of threading dislocations that etch to form nanopipes in the first porous layer, found through sub-surface BSE-SEM imaging \cite{Ghosh_Sarkar_Frentrup_Kappers_Oliver_2024}. Here, we reproduce the same effect, as Figure \ref{graphs}a demonstrates a clear increase in the ADD in the first porous layer with etching voltage. Tomography in this work can provide ADD through other layers of the stack. 

With Figure \ref{graphs}a, comparing the ADD obtained for each layer with their etching voltages, it is readily apparent that increasing etching voltage generally increases the density of active dislocations, regardless of the layer. More pertinent is that for the sample etched at 5\,V, there is a gentle decrease in active dislocation density moving downward through the stack, where deeper porous layers have fewer active dislocations than those close to the surface. This observation can account for the significant drop-off in porosity observed in deeper layers for samples etched at low voltages and the incomplete porosification of those layers. The sample etched at 8\,V remains relatively constant in the active dislocation density through the stack, and the sample etched at 10\,V shows an increasing trend.

It may be speculated that more dislocations are activated at higher etching voltages because the porous cells in doped layers become larger and less branched — pores are therefore more likely to intersect threading dislocations and, consequently, have the possibility of using them as etching pathways. For the etching of the first porous layer, all dislocations are initially exposed to the oxalic acid etchant at the material surface, so this effect alone cannot explain the resulting active dislocation density in the various layers of the samples due to the increased ADD with etching voltage there.

It should be noted that the native threading dislocation density local to the region of interest over which the tomograph was captured can vary significantly. A sample of the DBR template wafer, not having undergone electrochemical etching, underwent a silane treatment and was used to calculate the overall threading dislocation density across 20 regions of 1.5\,\textmu m $\times$ 1.5\,\textmu m \cite{OLIVER2006506}. Across the 20 regions, an arithmetic mean of the native threading dislocation density was found to be $2.55\times 10 ^{9}$\,cm$^{-2}$ with a standard deviation of $0.55\times 10 ^{9}$\,cm$^{-2}$. That the measured active dislocation density in the 10\,V sample is greater than this value could imply that nanopipes are being formed in defect-free regions of the NID GaN layer. However, the large active dislocation density observed for the 10\,V sample, in particular, can be attributed to the selection of the region of interest. Large porous fields in high voltage etched samples are associated with the etching of dislocation arrays rather than individual dislocations \cite{Ghosh_Sarkar_Frentrup_Kappers_Oliver_2024}. Dislocation arrays are small regions of very high local threading dislocation density. Hence, for the field of view of the tomograph to encompass one full porous field, the region selected needed to contain at least one dislocation array, and the region of interest on which the tomograph was ultimately captured actually features several of them. The areas in the AFM scans used to calculate the native density were randomly selected and subsequently included fewer arrays of threading dislocations. The selected regions for the 5\,V and 8\,V tomographs were also randomly chosen. We, therefore, propose that the field of view for this tomograph needed to be biased to a region of high local threading dislocation density for the experiment to succeed, not that the 10\,V tomograph has created nanopipes or etching pathways from non-dislocation regions.

Figure \ref{graphs}b probes the behaviour of individual dislocations and the number of layers in which they are active in the stack. For the sample etched at 5\,V, the decreasing trend here means that there are successively fewer dislocations with more active layers; indeed, there are very few dislocations active in four out of the five porous layers, and there are none active in all five, meaning that there are no dislocations exhibiting behaviour that could be described by the simple kebab model within the field of view of the tomograph. A successively decreasing trend of this kind could be attributed to a very simplistic model of etching behaviour. If each dislocation had a low constant probability of acting as an etching pathway in a given layer, independently of the previous etching of the dislocation or the etching of surrounding neighbours, one would expect a decrease in the number of dislocations with successively greater numbers of active porous layers. 

This straightforward decreasing trend concerning individual dislocations and the number of layers in which they are active in the stack is not observed, however, in the samples etched at 8\,V and 10\,V. Figure \ref{graphs}b also shows that the 8\,V etched sample follows the same initial trend of decreasing active dislocation density from one active layer through to four active layers, but then sees a dramatic increase in the active dislocation density for activity in all five layers. The same is true for the 10\,V etched sample, aside from a marginal increase in active dislocation density from 3 active layers to 4. This observation can no longer be attributed to an independent probability of a single dislocation etching in a given layer. Instead, the structures are dominated by dislocations that are either active throughout the whole stack or active for one layer only. For the three samples, the percentage of dislocations that form the 5-layer structures, and therefore behave according to the conventional kebab behaviour, are 0\,\%, 34.2\,\%, and 37.1\,\% for the 5\,V, 8\,V, and 10\,V samples, respectively. 

For each etching voltage, the most common number of active layers for a dislocation is 1. Such a dislocation cannot be said to form either a continuous or interrupted etching sequence and is thus distinguished as a `singlet' instead. An instance of this is highlighted in Figure \ref{confetti} in the grey box. These dislocations are activated once on the surface layer or in the sub-surface layers, producing one porous region and then deactivated. However, these singlets are not equally distributed among the different layers in the stack, and that distribution varies across etching voltages. The density of singlets across the porous layers for the three samples is given in Figure \ref{graphs}c. At 5\,V, the most likely position for a singlet is the first layer, where 37.9\,\% of singlets can be found. This high proportion is not seen in the 8\,V and 10\,V samples, where the first porous layer contains 3.0\,\% and 0\,\% of their respective singlet populations. Instead, the most likely positions for singlets are layers two and three for the 8\,V sample, which contains 26.9\,\% of the singlets, and layer five in the 10\,V sample, containing 38.6\,\% of singlets. 

Another aspect of dislocation etching behaviour to consider is the continuity of porous layer etching. Figure \ref{graphs}b describes the dislocations in terms of how many layers they are active in, but not which specific layers those are. For example, as outlined previously, a dislocation with a continuous etching sequence (e.g. etching layers two, three and four) and a separate dislocation with an interrupted etching sequence (e.g. etching layers two, three and five), are both included in the same data point in Figure \ref{graphs}b despite demonstrating different behaviour. From these tomographs, the percentage of dislocations with an interrupted sequence of porous layers in the 5\,V, 8\,V, and 10\,V samples is 26.9\,\%, 18.9\,\%, and 11.2\,\%, respectively, giving a decreasing trend with increasing etching voltage. 

The data presented here illustrates that our cascade model remains relevant across all samples. However, increasing the etching voltage increases the tendency to form `kebab' structures within the cascade networks, decreasing the likelihood of forming structures with interrupted etching sequences. At even higher etching voltages than 10\,V, these trends may be observed to continue. Also, at such voltages, it has been demonstrated that non-porous layers can undergo structural collapse during the etching, which will change the proliferation of etching significantly \cite{Ghosh_Sarkar_Frentrup_Kappers_Oliver_2024}. 

All of the above observations can be correlated to previous findings that show that in the porosification process, etching down only some dislocation pipelines penetrates deep into the doped/NID stack before other dislocation pipelines reach such a great depth \cite{massabuau_griffin_springbett_liu_kumar_zhu_oliver_2020}. It is possible that a particular dislocation type (edge, screw, or mixed) or a particular core structure has a higher etching rate, and this is an interesting area for further study, perhaps using correlative microscopy techniques. However, it is also possible that the onset of etching at a particular dislocation is probabilistic, with all dislocations having a fairly similar probability of etching but random chance, allowing some defects to etch to greater depths ahead of others. In either case, a dislocation that penetrates deep into the stack early in the process can then produce a pore that spreads widely in the lateral direction, not impeded by depletion zones from pores related to other defects. If such pores then intersect with a second dislocation, then etching may initiate at that second dislocation, producing a cascade and generating a depletion region which prevents the second dislocation from etching through from the overlying NID layer into the newly formed porous layer.

Whether all dislocations etch with equal probability or whether specific types of dislocations have different etch rates, changes to the probability of a dislocation etching will impact the likelihood of some dislocations penetrating deep into the stack before others do so. Both our results in this paper, and the previous work of Ghosh \textit{et al.} \cite{Ghosh_Sarkar_Frentrup_Kappers_Oliver_2024} suggest that this probability can be changed by varying the etching voltage, since the total active dislocation density has a strong dependence on etching voltage. We propose, therefore, that altering the etching voltage causes a change in the probabilities of dislocation etching, and also a change in the relative rates of the etching of dislocations compared to removal of material in the doped layers. The precise mechanism of how dislocation cores in NID GaN are able to etch at all is unclear; a change in the relative probability of etching of dislocations in NID GaN and n$^{+}$-GaN in doped layers reflects that the electrochemical mechanism of the two processes is different. It is possible that trap states associated with the defect are playing a role, where the production of free holes in the defect and production of free holes in defect-free doped material evolve differently as voltage is varied. It is also possible that silicon atoms segregate to threading dislocations, with diffusion then allowing for dislocations in NID layers to act as n-GaN with a low doping density. A probabilistic model might be built up which could predict the likelihood of morphologies dominated by kebabs or cascades and ascertain whether simple changes to the probabilities of the two processes described can reproduce the effects found here.

\section{Conclusions}

Serial-section tomography has captured the 3D structure of porous GaN DBRs etched by defect-driven mechanisms on a scale and resolution never demonstrated before. Detailed structural characterisation has previously been restricted to the first porous layer, but this work addresses the continuity and proliferation of porous morphologies throughout 5-layer structures. Tomographs, as outlined here, are not restricted to 5 layers, and the workflow developed is pertinent for porous DBR structures across many target wavelengths and indeed for a wide range of sub-surface porous multilayer structures, whether fabricated by defect-driven mechanisms or via the lithographic fabrication of deep trenches. Indeed, the utility of the approach applied is also relevant where porosification is achieved by methods other than electrochemical etching, such as thermal annealing. The fine voxel resolution employed, (2.0$\,\times$\,2.0\,$\times$\,5.1)\,nm in the best case, has permitted a mechanistic evaluation of the etching process for the first time, pioneering beyond the basic cross-sectional methods of pore morphology characterisation. The tomographs presented here offered the additional advantage of an extensive field of view, (2.7$\,\times$\,2.1$\,\times$\,0.5)\,\textmu m in the best case, meaning the behaviour of several hundred dislocations could be assessed, offering insight and generality beyond the probing of individual threading dislocations. 

The insights offered by volumetric reconstruction have allowed us to discover that rather than each dislocation forming a continuous uninterrupted pathway with an associated field of porosity in every doped layer, etching proliferates in a cascade, which can result in different dislocations switching on and off repeatedly in the course of the overall etching process. The extent to which complex cascades form is dependent on the etching voltage. We note that since the mechanism of turning on and off dislocations is suggested to depend on the relative probability of etching the dislocations compared to the doped layers, even where etching initially penetrates sub-surface via deep, lithographically defined trenches, there remains a finite probability of `turning on’ dislocation pathways, and this will influence the eventual morphology of DBR structures etched by that alternative approach. Broader efforts to optimise the fabrication of porous GaN DBRs, fabricated by either a defect-driven or lithography-assisted electrochemical etching process, can therefore benefit from both the conclusions drawn and the methodologies outlined in this work.

\titleformat{\subsubsection}{\normalfont\bfseries}{}{0pt}{}

\subsubsection{Acknowledgements}

This work was supported by the Ernest Oppenheimer Trust at the University of Cambridge and by the Royal Academy of Engineering under the Chairs in Emerging Technologies scheme, funded by the Department for Science, Innovation and Technology (DSIT). We acknowledge the support of the Wolfson Electron Microscopy Suite and the use of the Zeiss Crossbeam 540 funded by Royce under grant EP/R008779/1. The EPSRC also supported this research under Grant Nos. EP/R03480X/1, EP/W03557X/1, EP/X015300/1, EP/N509620/1, and EP/R513180/1 as well as under Project Reference 2278538. We also acknowledge funding from The Armourers and Brasiers’ Gauntlet Trust.

\newpage

\subsubsection{Supplementary information available}
See the supplementary material for a comparison of sub-surface BSE-SEM imaging and reconstructed plan-view images as derived from serial-section FIB-SEM tomography of the first porous layer for the porous GaN-on-Si DBRs etched at 5\,V (Supplementary Figure 1) and 8\,V (Supplementary Figure 2). We also provide orthographic perspective 3D renders derived from serial-section FIB-SEM tomography, depicting the pore morphology of 5-pair porous GaN-on-Si DBRs etched at  5\,V and 8\,V after image segmentation via intensity thresholding (Supplementary Figure 3).

\newpage

\bibliographystyle{elsarticle-num} 
\bibliography{reference.bib}



%\begin{thebibliography}{00}

%% For numbered reference style

%\bibitem{lamport94}

%\end{thebibliography}

\end{document}